\documentclass[a4,preprint]{revtex4}
\usepackage{graphicx}
\begin{document}
\author{V.M. Shilov}
\affiliation{Bogoliubov Lab. Theor. Phys., 141980, JINR, Dubna,
Russia}
\title{Sub-barrier fusion of intermediate and heavy nuclear systems}

\begin{abstract}

A potential  model to describe the total  cross section
of nuclear fusion reactions  at low energies is proposed.
It is shown that  within an approach with a  simple, single barrier
potential a satisfactory description of data is hindered, while with
models with  more complex interactions, e.g., with two-barrier
potentials, a good description of data can be achieved in a large interval
of colliding nuclei and center of mass energies.
In particular, the two-barrier model allows to describe data at low,
sub-barrier energies where data exhibits a steep falloff
of the cross section. It is also shown that the
position of the two barriers is almost independent up on masses of
colliding nuclei. Comparison with available experimental data is
presented as well.

\end{abstract}
\maketitle
\section{Introduction}
 Investigation of fusion reactions at low center of mass
 energies ($E\simeq 30-100 MeV$) is of interests
 not only in context of study of the  mechanism
 of nucleus-nucleus interaction, but also  supplies unique
 information on astrophysical processes  on stars, in particular
 on the so-called astrophysical $S$-factor, which is directly related
 with the e.g., flux of solar neutrinos. Experimental measurements
 at energies well above the Coulumb barrier
 started a few decades ago~\cite{GGLT74,GM74} and shown that the cross section
 exhibits a smooth behaviour as a function of the center of mass  energy.
However, measurements of very small cross sections ($\sigma_{ER}\approx 10^{-2} mb$),
near and below the Coulomb
barrier~\cite{JRJ04,KSH86,JER02,OIK96,MBD99,JBE06,DHD07}
have shown an unexpected  steep falloff of the cross section, named
as "hindrance" effect.
Theoretical efforts to describe such a shape
  generated a number of models~\cite{S98,HRD03,DP03,JEB04,ME06,IHI07,DHD07}
within which different reaction mechanisms have been suggested
to reconcile data  at extreme sub-barrier energies.
Since different models  describe data equally well,  in order to distinguish
the proposed  mechanisms one needs more measurements at
even lower energies; a comparison with model  predictions will allow to enlighten the
peculiarities of the
reaction mechanism.

In the present paper a further development
of the so-called  "critical distance model"~\cite{GGLT74,GM74} is presented.
Remaind, that within such a model, in order to create the compound nucleus,
 the colliding nuclei have to approach some "critical
distance"  $R_{cr}$ with $R_{cr}=r_{cr}^0(A_1^{1/3}+A_2^{1/3})$
($r_{cr}^0 \approx 1 fm$).
At such distances the colliding nuclei loss their individual structure forming
a single highly excited system.
Since at the initial stage the  collision is mainly superficial
the deformation nuclei can be neglected and, correspondingly,
the ion-ion potential can be  calculated within the so-called
"frozen approximation"~\cite{GGLT74,GM74}.
It should be noted that when the
 density in the overlap region becomes larger than equilibrium density,
 the Pauli principle forbids  further penetration of nuclei in to each other.
 This gives rise of  a repulsion part in the
 ion-ion potential~\cite{NTB75,BRST77} at such critical distances.
  At smaller distances $r < R_{cr}$ the colliding nuclei forming the compound system
  become deformed so that the densities cannot be considered  "frozen" anymore.
  This leads to a  drastic attenuation  of  the potential up to
  values $Q_R=(M_{CN}-M_1 - M_2)\,c^2$, where $Q_R$ is the threshold energy
 of the reaction and $M$ denotes the corresponding mass of the nuclear system.
 It is   clear, that if the critical distance $R_{cr}$
 is smaller than  typical positions of the Coulomb barrier, this
 results in a two barrier potential; besides the Coulomb barrier here
 appears another, inner barrier $V_C(R_{cr})$. The Coulomb barrier dominates at small values of
 atomic numbers $Z$, however with increasing of $Z$
 the first, inner  barrier becomes similar and even larger than the Coulomb barrier.
Usually the strength of  potentials $V_C(R_{cr})$ is characterized
 by a parameter $\eta_{cr}$
($\eta = Z_1\,Z_2/(A_1^{1/3}A_2^{1/3})$)~\cite{Sch86} which demonstrates
 that already at
$\eta_{cr} > 71.3 $ the first barrier, $V_C(R_{cr})$,  becomes larger than
the  Coulomb one, $V_B(R_B)$.
For light and intermediate nuclei
$V_C(R_{cr})$ is always smaller than
  $V_B(R_B)$   and it
contributes only  at large angular momenta and  high energies.

The aim of this work is to investigate the effects of the two barriers
in the ion-ion potential
and   to explain the observed hindrance as effects of
 resonance tunneling
through such a two-humped potential~\cite{S98}.
It will be  shown that the inner barrier manifests  also in deep sub-barrier
cross sections of medium magic and semi-magic nuclei even at
 $\eta < \eta_{cr}$.
Particular attention is paid to fusion reactions of nuclei with mass
number closed to
the magic ones. In this case according to the nuclear shell model
the single particle energy levels are closed and the nucleus acts as a hard
sphere impeding the deformation during the fusion process. As a result,
the tunneling of undeformed ions occurs without forming a compound system and
the interference between two barriers can become important.
This effect is studied in the present paper in details.

  Since the procedure of finding the  exact solution of the
  Schr\"odinger equation (actually, this is a system of
  coupled equations)
  is a cumbersome and rather  tedious task, in the present approach
  a  semiclassical approximation is adopted  within which
  the  system is considered   uncoupled  and each equation solved separately.
  The resulting solution contains a number of free parameters which determine the general
  form of the inner potential. These parameters are to be found from a
  combined fit of the experimental data for different pairs of colliding nuclei.
  The possibility to apply the proposed model to other kind
  of  processes with heavy ions is discussed as well.

\section{Formalism}

The equations for radial wave functions
$R_{\alpha L_\alpha}(r)$, describing the relative motion of the two ions
in the   coupled channels method  can be written as ~\cite{DLW83}

\begin{eqnarray}
\left[ \frac{d^2} {dr^2} + \frac{2 \mu} {\hbar^2}(E-\varepsilon_\alpha -
V_\alpha (r)) -\frac{L_\alpha (L_\alpha +1)} {r^2} \right]
R_{\alpha L_\alpha}(r) = \frac{2 \mu} {\hbar^2} \sum_\beta V_{\alpha\beta}(r)
R_{\beta L_\beta}(r).
\label{a2}
\end{eqnarray}

Here $L_\alpha$ is the  orbital angular moment, $V_\alpha(r)$ is the diagonal part
of ion-ion potential in the $\alpha$ channel with excitation energy
$\varepsilon_\alpha$ and
$V_{\alpha \beta}(r)$ is the  coupling potential between  channels.
Eventually, $\mu$ is the reduced mass of the system.

 As already mentioned, the coupled system (\ref{a2})
 can be decoupled and solved
 by considering the semiclassical approximation~\cite{LR84}.
 Within this approximation the system is spinless for both channels
 and the energy levels are considered  as a  two-level degenerated
 system in the ground state.
 Introducing the notation  $\chi_L^{(\pm)}(r)=R_{0L}(r)\pm R_{1L}(r)$ and
 using the equalities
 $ V_0 (r)=V_1 (r)$ and
$ V_{01} (r)=V_{10} (r)$, one obtains
\begin{eqnarray}
\left[ \frac{\hbar^2}{2 \mu} \frac{d^2} {dr^2} + E-
\frac{L(L+1) \hbar^2}{2 \mu r^2} - (V_0(r) \pm V_{01}(r)) \right]
\chi_L^{(\pm)}(r)=0.
\label{a8}
\end{eqnarray}
In the simplest case when
the fusion process is supposed to be defied as a conserved flux of ions
tunneling throughout the  Coulomb barrier the system can be solved analytically
resulting in the known  Wong formulae~\cite{Won73}. Then in this case
the  fusion cross section can be expressed as
\begin{eqnarray}
\sigma_{fus}(E) = \frac{ \hbar \omega R_B^2}{4E} \left\{
\ln [1+ \exp \frac{2 \pi}{ \hbar \omega} (E-V_0 (R_B)-V_{01}(R_B)) ]\right\} + \nonumber \\
\frac{ \hbar \omega R_B^2}{4E} \left\{ \ln [1+ \exp \frac{2 \pi}{ \hbar \omega} (E-V_0 (R_B)+V_{01} (R_B)) ] \right\}.
\label{a9}
\end{eqnarray}
In (\ref{a9})  the radii  in both channels have been taken the same.

In solving the equation (\ref{a8}) the quasi-classical method
Wentzel–Kramers–Brillouin (WKB), originally
proposed to describe fission reaction~\cite{MS78,BA89},
has been used. Accordingly, the
 penetrability $T$ can be written
\begin{eqnarray}
T=T_C T_B/(\{1+[(1-T_C)(1-T_B)]^{1/2} \}^2 \cos^2\nu
+\{1-[(1-T_C)(1-T_B)]^{1/2} \}^2 \sin^2 \nu),
\label{a11}
\end{eqnarray}
where for the sake of brevity in the above equation the
angular momentum $L$ dependence has been suppressed;
the  subscripts  $B$ and  $C$ stand for the  Coulomb and inner barriers, respectively.
The corresponding penetrability  $T_{C,B}$
within the  WKB-approximation at sub-barrier energies is
\begin{equation}
 T_{C,B} \, (E) = \left\{ 1+ \exp \,[2 \, S_{C,B}(E) \,] \right\} ^{-1},
\label{a12}
\end{equation}
\noindent
where the  actions $S_{C,B}(E)$
and phases $\nu$
are defined by integrals over the interval between corresponding return points
in ion-ion potentials $V_L(r)$
\begin{equation}
 S_{C,B}(E) = \int \left\vert 2\,\mu \,[ V_L(r)-E ]/\hbar^2
 \right\vert ^{1/2}dr
\label{a13}
\end{equation}
\begin{equation}
 \nu = \int \left\vert 2\,\mu \,[ E-V_L(r) ]/\hbar^2
 \right\vert ^{1/2}dr.
\label{a14}
\end{equation}

At larger penetrability,  $T_{C,B} > 0.45$,    the tunneling
probabilities can be  parametrized analytically by known
Hill-Wheeler formulae with near-barrier parameters of curvatures

\begin{equation}
 \hbar \omega _{C,B} =\left\vert \frac{ \hbar^2}{\mu} \frac{d^2 \,V_{C,B}(r)}{dr^2}
 \,\right\vert^{1/2}_{r=R_{C,B}}.
\label{a15}
\end{equation}

\begin{eqnarray}
 T_{C,B}(E)= \left\{
 1+\exp \left[ \frac{2\pi}{ \hbar \omega_{C,B}}
  (E-V_{C,B}) \right]
 \right\}^{-1}.
\label{a16}
\end{eqnarray}
Here $V_{C,B}$, $ R_{C,B}$ are potentials and radii of inner and Coulomb barriers of
each partial wave.

Then the   averaged, over  $\nu$ in eq. (\ref{a11}),
penetrability can be written as
\begin{eqnarray}
T=T_C T_B/[1-(1-T_C)(1-T_B)].
\label{a17}
\end{eqnarray}
Exactly the same expression for the penetrability
can be obtained by summing ingoing and
outgoing fluxes in the interval  between the two barriers.
Note, that by averaging in obtaining    (\ref{a17}) all the interference effects
have been lost. So that a comparison of results by unaveraged
 formula (\ref{a11}) and
by equation (\ref{a17}) will reflect the role of
interference effects in reactions of heavy ion fusion
at low energies.

\section {Data analysis}

We have analyzed fusion reaction for a variety of combinations
of colliding ions available for experimental studies.
The corresponding reactions together with some their specific
characteristics ($Q$-reaction, minimal energy, Bass potential and
angular momentum  of the evaporation residue) are listed in Table~\ref{tab1}.
In our approach  these quantities serve as input parameters.
Our analysis consists on changing the form of the
of ion-ion potential and computing the fusion cross section with further
fit of the potential parameters to obtain a best description of data.
As the outer part of the Coulomb potential the
Wood-Saxon form with parameters $V, R, a$ has been used.
The coupled channel potential has been taken in the form

\begin{eqnarray}
V_{01}(r) = \frac{\beta_\alpha R_0}{\sqrt{4\pi}} \left[
-\frac{dV_0(r)}{dr}+\frac{3}{2 \lambda+1}Z_1Z_2e^2
R_0^{\lambda-1}/r^{\lambda+1} \right],
\label{b1}
\end{eqnarray}

\begin{table}[!th]
\caption{The considered reactions and the relevant
parameters used in the numerical calculations. $Q_R$ is the threshold
energy of the reaction, $E_{min}$
is the minimal energy at which the experimental measurements
have been performed, $J_{ER}$ is the parameter used to distinguish
the two deexcitation process, eq. (\ref{b5}), $V_{Bass}$
is the Bass potential used in most phenomenological calculations. }
\begin{tabular}{|c|c|c|c|c|c|}
\hline
 Reactions   &$-Q_R(MeV)$ & $V_{Bass}(MeV)$ & $E_{min}(MeV)$
 &$J_{ER}$ & $Refs.$   \\
\hline
$^{64}Ni+^{64}\!Ni$  & 48.71  &  95.7 & 85.55  &  -   &\cite{JRJ04}\\
$^{60}Ni+^{89}\!Y $  & 90.64  & 129.9 & 121.4  &  -   &\cite{JER02}\\
$^{90}Zr+^{92}\!Zr$  &154.0   & 180.9 & 169.6  & 17.0 &\cite{KSH86}\\
$^{28}Si+^{64}\!Ni$  &  1.78  &  52.3 &  44.0  &  -   &\cite{JBE06}\\
$^{16}O+^{208}\!Pb$  & 46.49  &  76.0 &  62.7  &  6.0 &\cite{OIK96,MBD99}\\
\hline
\end{tabular}
\label{tab1}
\end{table}
\noindent
where $\lambda$ is the  multipolarity of the level
and  the  dynamical deformation parameter $\beta_\alpha$
is taken from independent experiments.
In the present calculations
the  quantity  $(\beta_\alpha R_0)/\sqrt{4\pi}$ has been taken
constant and the same for
 all reactions listed  in Table \ref{tab1}
 ($\beta_\alpha R_0/\sqrt{4\pi}=0.3$).

 The form of the inner part of the central potential
imitating the repulsive interaction
is determined, in the present paper,  by the  expression

\begin{eqnarray}
 \Delta V_0(r) = V_1 \; (R_m-r) ^2.
\label{b2}
\end{eqnarray}
which, at
$r<R_m$ ($R_m$ is determined by the position of the minimum of the
central potential) is similar to one used in Ref.~\cite{NTB75}.
To compensate the increase  of the potential (\ref{b2}) at low distances
(increase of $R_m-r$), an
additional  damping factor
\begin{eqnarray}
  g(r) = \left(  1 + \exp \frac{r-R_1}{ \Delta} \right) ^{-1}
\label{b3}
\end{eqnarray}
has been introduced with parameters $R_1$ и $\Delta$ as free ones.
In such a way the total  potential results in a continue function
of the distance $r$ with continue first derivatives, as it should be for
a nuclear potential. Explicitly, the potential reads as

\begin{eqnarray}
 \widetilde V_0 (r) \pm \widetilde V_{01}(r) = Q_R \cdot g(r) +
(V_0(r) \pm V_{01}(r)) \cdot (1 - g(r)).
\label{b4}
\end{eqnarray}

\begin{table}
 \caption{The final values of the phenomenological parameters of the model obtained by
 fitting the experimental data. The quantities $V$, $R$ and $r_0$ are the
 magnitude of the barrier, its position and nuclear
  parameter for Coulomb and inner
 potentials, respectively. }
\begin{center}
\begin{tabular}{|c|c|c|c|c|c|c|c|c|}
\hline
 Reactions &$V_B$ &$R_B$ &$r_0^B$ & $\hbar \omega_B$&$V_C$
 &$R_C$ &$r_0^C$& $\hbar \omega_C$  \\
\hline \
$^{64}Ni+^{64}\!Ni$   & 93.55 & 11.28&  1.41  & 3.57   & 85.11 & 6.68 & 0.84 & 12.17 \\
$^{60}Ni+^{89}\!Y $   &129.65 & 11.50&  1.37  & 4.33   &122.33 & 6.92 & 0.83 &  9.76 \\
$^{90}Zr+^{92}\!Zr$   &173.63 & 12.35&  1.37  & 3.46   &178.23 & 7.87 & 0.87 & 10.56 \\
$^{28}Si+^{64}\!Ni$   & 51.26 & 10.35&  1.47  & 3.92   & 45.05 & 6.04 & 0.86 & 19.7  \\
$^{16}O+^{208}\!Pb$   & 74.47 & 11.98&  1.42  & 4.76   & 61.86 & 7.44 & 0.88 & 14.83 \\
\hline
\end{tabular}
\end{center}
\label{totl}
\end{table}
The potential (\ref{b4}) is used to calculate the nuclear penetrability $T$, eq.
(\ref{a11}). Besides penetrability, the fusion cross section is
determined by the probability of deexcitation of the compound system
via  fission and evaporation processes.
In a compound system formed by light nuclei the
evaporation process prevails, while for heavy components the
fission reaction dominates. In the medium region both processes
are of an  equal importance. To distinguish between these two processes
of deexcitation we use
a simplest model, suggested in \cite{Mat84}. It consists in merely
 multiplication of
 the penetration probabilities in(\ref{a11}) and (\ref{a16})
by the  "deexcitation" factor
\begin{eqnarray}
  W_{ER}(L) = \left(1 + \exp \frac{L-J_{ER}}{2}\right) ^{-1},
\label{b5}
\end{eqnarray}
\noindent
where angular momentum $J_{ER}$ separates fission and evaporation residue
cross sections, cf. Table~\ref{tab1}.

Having determined the form of the cross section we fitted the
free parameters to obtain a good description of data in a large region of energies and
nuclear masses.
The results of our analysis are shown in Table~\ref{totl} where
the fitted values of the relevant parameters of the model are listed.
The main result from the Table can be formulated as follow:
 since for all combinations of the colliding ions the difference
  $ R_B-R_C$ is almost constant ($ \approx 4.5 fm$) the distnace between two barriers
  is approximatively the same for all the considered reactions.
  Moreover, as expected $ R_B-R_C$  is roughly twice  the diffuse edge of nuclei.

\begin{figure}[ht]      
\includegraphics[width=0.48\textwidth,angle=0]{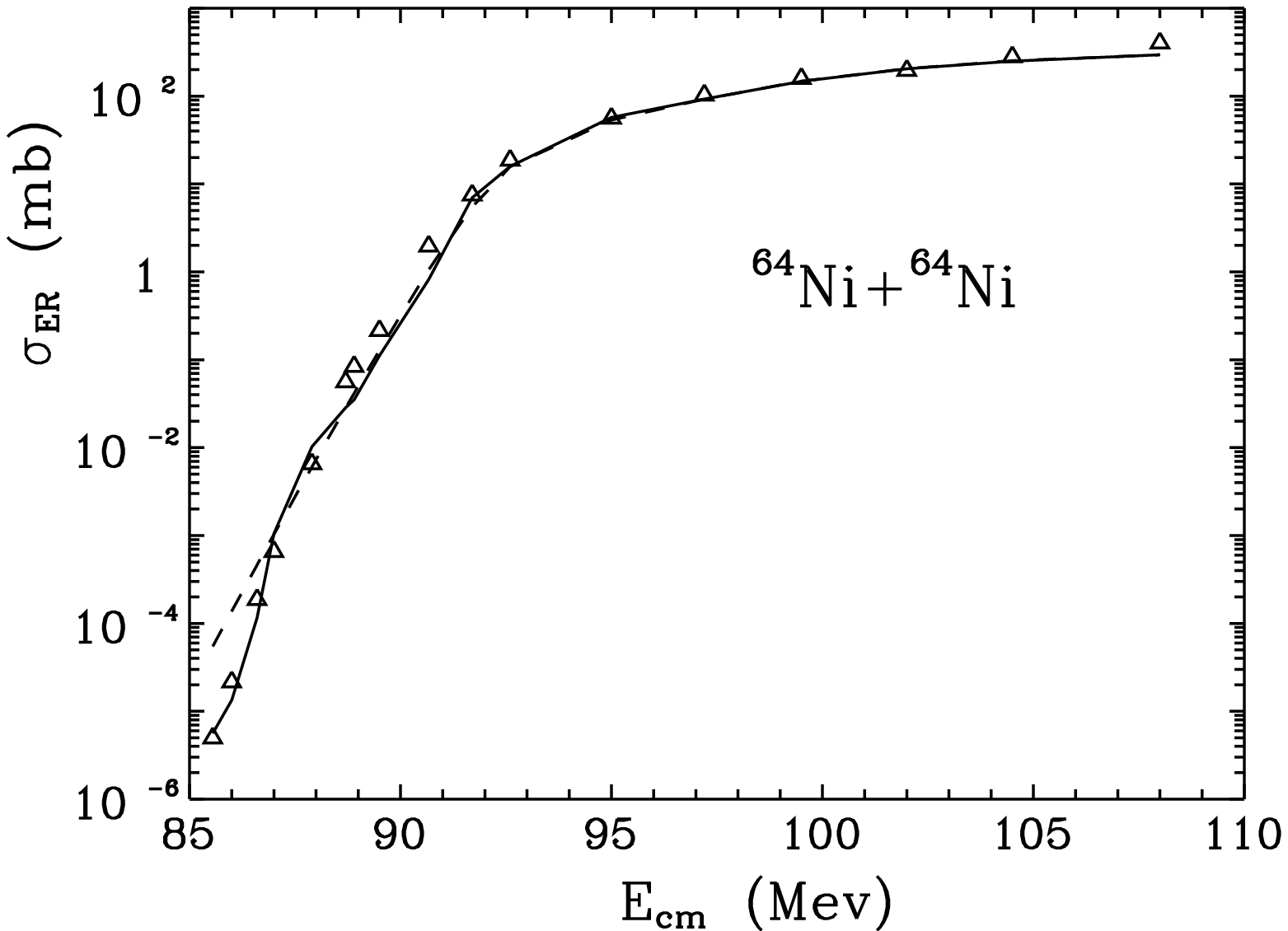}
\includegraphics[width=.48\textwidth,angle=0]{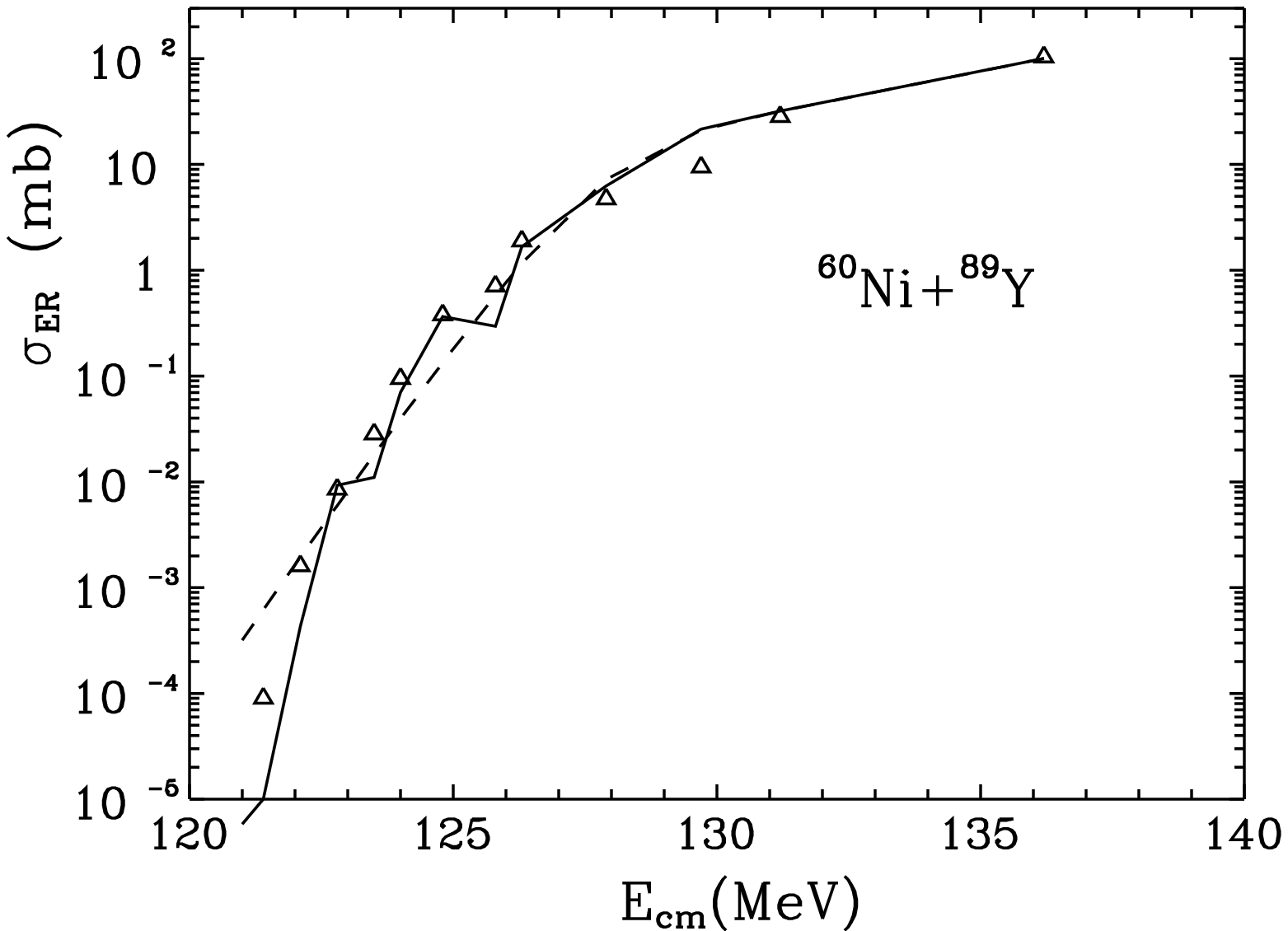}
\caption{The calculated cross sections Eq. (\ref{a9})
 for the reactions
 $^{64}Ni+^{64}Ni$, left panel and for the reaction
 $^{60}Ni+^{89}Y$, right panel.
 Solid curves correspond to full calculations with including
 the interference effects in penetrability, eq. (\ref{a11});
 dashed curves are the results of calculations without interferences,
 eq. (\ref{a17}). Experimental data correspond to the measured
 evaporation residues~\cite{JRJ04,JER02}.
}
  \label{fig1}
\end{figure}

As an example, in Figs.~\ref{fig1} we present results of
 the cross section calculated (solid lines)
   within our model with parameters listed
in Table ~\ref{totl} for two combinations of colliding nuclei.
Since in such kind of reactions the colliding nuclei are considered as light,
the theoretical cross section is compared with the experimentally measured
evaporation residue cross sections.
It can be seen that within two-barrier model a good description of data can be
achieved in the whole interval of considered energies for both combinations of colliding ions.
It should be pointed out that the hindrance effect has been obtained only
by an accurate account of the interference effects, as above discussed.
Without interference, i.e. cross section by formula (\ref{a17})
(dashed lines in Figs.~\ref{fig1}) an increase of the  slope
of the cross section at low energies cannot be obtained.
Our results are in an agreement with  the ones of Ref.~\cite{TSS91},
 the  fusion cross sections for the reactions
$^{96}\!Zr + ^{124}\!Sn$,
$^{86}\!Kr + ^{123}\!Sb$ и $^{50}\!Ti + ^{208}\!Pb$, where
$\eta > \eta_{cr}$
have been analyzed.

\begin{figure}[!bht] 
\includegraphics[width=.5\textwidth,angle=0]{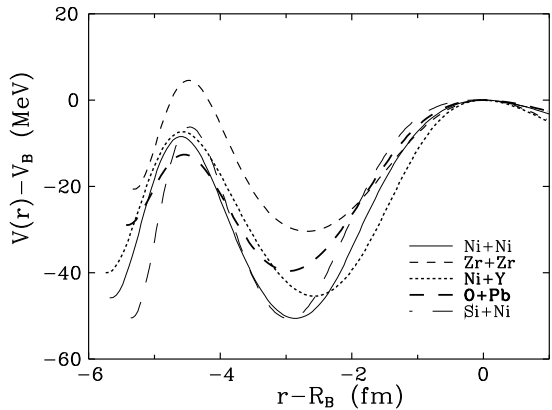}
\caption{The phenomenological potential, eq. (\ref{b2}),
as a function of the "scaling" variable $r-R_B$ for a set of colliding ions.
The scaling-like behaviour of the inner potential is clearly seen at
$r-R_B\simeq -4.5 fm $.    }
\label{fig3}
\end{figure}

As an illustration of the obtained potential in Fig.~\ref{fig3}
 we plot the fusion potential without the Coulomb barrier,
 $V(r)-V_B$ as a function
of the distance from the Coulomb barrier $r-R_B$ for five combinations of the
colliding particles. It is clearly seen that the position of the barriers
relative to $R_B$ is roughly the same. This is an important result
since if so, predictions for other sets of colliding nuclei can be easily
performed.

 Another important result obtained during the phenomenological fit
 is that at $E=Q_R$ the phase $\nu$ in eq. (\ref{a14})
 becomes   $\approx\pi /2$, which is a clear evidence of the
 fact that in the inverse to the fusion channel
   the decaying system was in its
first quasi-stationary state. It means that the obtained potential (\ref{b4})
can be successfully applied in describing the decay of heavy nuclei into
heavy clusters.

\section{Conclusion}
The fusion processes of two nuclei have been analyzed
within a model with complex shaped  ion-ion potential.
It has been shown that the observed hindrance effect in the
fusion cross section at extremely low energies can be
described by a two-barrier potential. The corresponding shape and values of the
phenomenological parameters have been obtained by a fit of experimental
data and it has been shown that the  potential depends rather on
the relative distance from the Coulomb barrier that the distance between the
colliding ions. It has been found that the interference contribution in to
the penetrability $T$ plays a crucial role in understanding the drastic change
of the slope of the cross section at extremely sub-barrier energies.
The obtained phenomenological potential
possesses a scaling-like behaviour, in sense that it is almost the same
for different ions at the same values of the distance from the Coulomb
barrier. This allows to apply our model to other kind of reactions, e.g. the heavy
cluster decay processes.

\newpage


\end{document}